\documentclass[aps,prl,superscriptaddress,twocolumn,showpacs]{revtex4}

\usepackage{bm}
\usepackage{graphicx}
\usepackage{amsfonts}
\usepackage{amsmath}

\begin{document}

\title{Detection of XY behaviour in weakly anisotropic 
quantum antiferromagnets on the square lattice}

\author{Alessandro Cuccoli}
\affiliation{Dipartimento di Fisica, Universit\`a di Firenze,
    Via G. Sansone 1, I-50019 Sesto Fiorentino, Italy}
\affiliation{Istituto Nazionale per la Fisica della Materia, 
UdR Firenze, Via G. Sansone 1, I-50019 Sesto Fiorentino, Italy}    
\author{Tommaso Roscilde}
\altaffiliation{Current address: Department of Physics and Astronomy,
University of Southern California, Los Angeles, CA 90089-0484.}
\affiliation{Dipartimento di Fisica, Universit\`a di Firenze,
    Via G. Sansone 1, I-50019 Sesto Fiorentino, Italy}
\affiliation{Istituto Nazionale per la Fisica della 
Materia, UdR 
Firenze, Via G. Sansone 1, I-50019 Sesto Fiorentino, Italy}    
\author{Ruggero Vaia}
\affiliation{Istituto di Fisica Applicata 'N. Carrara', Consiglio
Nazionale delle Ricerche,
Via Panciatichi~56/30, I-50127 Firenze, Italy}
\affiliation{Istituto Nazionale per la Fisica della Materia,
UdR Firenze, Via G. Sansone 1, I-50019 Sesto Fiorentino, Italy}    
\author{Paola Verrucchi}
\affiliation{Dipartimento di Fisica, Universit\`a di Firenze,
    Via G. Sansone 1, I-50019 Sesto Fiorentino, Italy}
\affiliation{Istituto Nazionale per la Fisica della Materia,
UdR Firenze, Via G. Sansone 1, I-50019 Sesto Fiorentino, Italy}    
\date{\today}

\begin{abstract}
We consider the Heisenberg antiferromagnet on the square 
lattice with $S=1/2$ and very weak easy-plane exchange 
anisotropy; by means of the quantum Monte Carlo method, based 
on the continuous-time loop algorithm, we 
find that the thermodynamics of the model 
is highly sensitive to the presence of tiny anisotropies 
and is characterized by 
a crossover between isotropic and planar 
behaviour. We discuss the mechanism underlying the crossover 
phenomenon and show that it occurs at a 
temperature which is characteristic of the model.
The expected Berezinskii-Kosterlitz-Thouless transition 
is observed below the crossover: a finite range 
of temperatures consequently opens for experimental 
detection of non-critical 2D XY behaviour.
Direct comparison is made with uniform susceptibility data 
relative to the $S=1/2$ layered antiferromagnet Sr$_2$CuO$_2$Cl$_2$.

\end{abstract}

\pacs{75.10.Jm, 05.30.-d, 75.40.-s, 75.40.Cx}

\maketitle

Predictions of the  Berezinskii-Kosterlitz-Thouless (BKT) 
theory~\cite{BKT}
for topological ordering with zero order parameter
have been verified in many real systems, such as 
superfluid or superconducting films~\cite{Mooij94} 
and Josephson junction arrays~\cite{FaziodZ01}. 
However, despite the BKT theory 
being originally formulated as referred to 2D planar magnets, 
evidences of XY behaviour in real magnets are 
weak and limited to very peculiar cases~\cite{BKTexp}.
On the other hand, for $S=1/2$ there exist several cuprous 
oxides that, besides being excellent realizations of
2D antiferromagnets,
are known~\cite{Johnston97} to display a weak 
exchange easy-plane (EP) anisotropy:
unambiguous observations of XY 
critical behaviour in such systems are not easy to achieve,  
due to both the weakness of the anisotropy 
and the existence of finite inter-layer coupling.
As for the former point, the EP anisotropy observed 
in real magnets is 
usually $10^{-4}\div10^{-3}$ times the isotropic 
exchange coupling, and the signatures of BKT critical 
behaviour are 
often either too weak to be extracted from the isotropic 
thermodynamics or too close to the critical temperature to be 
experimentally accessible.
The residual inter-layer coupling, even if
orders of magnitudes smaller than the intra-layer one,
drives the system towards a 3D transition, which is
actually triggered by the divergence of 2D 
intra-layer spin correlations. Therefore
purely 2D critical behaviour of diverging quantities is most 
often masked by the onset of 3D long-range order.

In this work  we show that several non-diverging quantities are 
sensitive to the presence of EP anisotropy and display an 
evident and detectable crossover between 
isotropic and XY behaviour above 
the expected BKT transition. Such crossover occurs at 
a temperature which is characteristic of the 
model and is marked by peculiar features 
in the temperature dependence of non-critical observables.
In particular, we present quantum Monte Carlo
(QMC) data for the uniform susceptibility, the 
finite-size staggered out-of-plane magnetization, the 
specific heat and the density of in-plane vortices.

We consider the $S=1/2$ easy-plane antiferromagnet on the square 
lattice described by the Hamiltonian
\begin{equation}
 \hat{\cal H} = \frac{J}{2} \sum_{{\bm i},{\bm d}}
 \left( \hat{S}_{\bm i}^x\hat{S}_{{\bm i}+{\bm d}}^x+ 
 \hat{S}_{\bm i}^y\hat{S}_{{\bm i}+{\bm d}}^y + 
(1-\Delta)\hat{S}_{\bm i}^z\hat{S}_{{\bm i}+{\bm d}}^z\right)
 \label{e.xxzmodel}
 \end{equation} 
 where ${\bm i}=(i_1,i_2)$ runs over the sites of a square lattice,
 ${\bm d}$ connects each site to its four nearest neighbours,
 $J >0$ is the antiferromagnetic exchange coupling and
 $\Delta \in (0,1]$ is the EP anisotropy parameter. We 
 use the reduced temperature $t=T/J$. 
 
The above model is studied by means of 
QMC simulations based on the continuous-time loop algorithm 
~\cite{Beard,Evertz02,CRTVVprb02} for 
$\Delta$ = 0.001, 0.02, and 1, and on lattice sizes 
from $L = 64$ to $L = 200$. Each MC run consists of $10^4$
thermalization steps, followed by $1\div1.5 \times 10^5$
MC steps for measurements. Whenever possible,
we implement improved estimators~\cite{Evertz02}
for the quantities of interest. 

The class of EP antiferromagnets described by 
Eq.~(\ref{e.xxzmodel}) is characterized by 
the possibility for the model to support non-linear 
topological 
excitations in the form of in-plane vortices (V)  
and antivortices (AV). 
Such excitations play a fundamental role in determining the 
thermodynamics and critical behaviour of the system, as 
described by the BKT theory relative to the classical planar
rotator model. When both quantum and 
out-of-plane fluctuations are present,  
the BKT behaviour persists
even in strongly quantum ($S=1/2$) nearly isotropic 
($\Delta\simeq 10^{-3}$) models~\cite{CRTVVprb02}.

Nevertheless, there exist significant differences 
between the standard BKT phenomenology
and the XY behaviour actually observed in quantum nearly 
isotropic antiferromagnets, where the out-of-plane spin 
component plays an essential role above the transition.

Let us consider a strongly anisotropic EP model 
($\Delta\lesssim 1$): 
At high temperatures the system, despite being
disordered,  
already supports topological excitations in the form of
well separated V and AV in the easy plane.
As $t$ decreases V and AV attract each other, 
till V-AV pairs begin to  
form and a maximum in the specific heat is correspondingly
observed. At the critical temperature $t_{_{\rm BKT}}$ 
V and AV are all paired and the topological transition 
occurs.

In the nearly isotropic case ($\Delta \ll 1$), the picture is 
modified as follows.
At high temperature the system is isotropically disordered.
When $t$ decreases
the EP anisotropy becomes effective enough to stabilize planar 
configurations and allow for V and AV to appear in the 
easy plane. This phenomenon, occurring at a temperature  
hereafter indicated as $t_{\rm co}$,  may be thought of as a 
crossover between the isotropic and a more genuine XY 
behaviour, 
and suggests the temperature range $t_{_{\rm BKT}}<t<t_{\rm co}$
as the most appropriate for experimental observations.

\begin{figure}
\includegraphics[bbllx=26pt,bblly=-10pt,bburx=516pt,bbury=500pt,%
     width=67mm,angle=0]{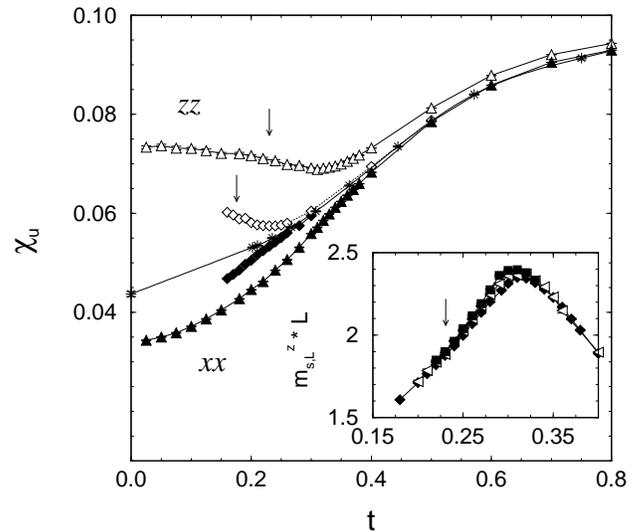}
 \caption{\label{f.chiU}
Uniform susceptibility of the EP model for 
$\Delta=0.02$ (triangles) and  $\Delta=0.001$ 
(diamonds). Open (full) symbols represent the out-of-plane (in-plane)
susceptibility.  The stars represent the 
uniform susceptibility in the
isotropic case~\cite{Beard,KimT98,MakivicD91}.
Inset: finite-size magnetization 
of the EP model for $\Delta = 0.02$ and 
$L=64$ (diamonds), 96 (left triangles), and 128 (squares).
Error bars are smaller than symbol sizes;
arrows are the critical temperatures as estimated
via finite-size scaling~\cite{CRTVVprb02}.}
\end{figure}

We consider the dimensionless uniform susceptibility 
$\chi^{\alpha\alpha}_{\rm u}$, defined as
\begin{equation}
\chi^{\alpha\alpha}_{\rm u} =
\frac{J}{L^2} \sum_{\bm i \bm j}  
\int_0^\beta d\tau ~ 
\left\langle \hat{S}^{\alpha}_{\bm i}(0) 
\hat{S}^{\alpha}_{\bm j}(\tau) 
\right\rangle~~.
\end{equation}
Fig.~\ref{f.chiU} shows its temperature dependence,
which we find strongly characterized by the appearence
of a minimum in the $zz$ component. 
A similar minimum has been
indeed experimentally observed in Sr$_2$CuO$_2$Cl$_2$ 
and suggested to be related to the onset of 2D XY 
behaviour~\cite{Vakninetal97}. 

A sound argument in favour of this interpretation is the 
following. An infinitesimal uniform magnetic field
causes a finite response  in an antiferromagnet via the 
canting mechanism: adjacent spins antialigned 
in the plane perpendicular to the field cant out of such
plane and give rise to a net magnetization, 
while a negligible response comes from 
spins anti-aligned along the field direction. 
On this basis, we interpret Fig.~\ref{f.chiU}.
At high temperature, $\chi_{\rm u}^{zz}$ 
and $\chi_{\rm u}^{xx}$ behave in the same way 
and decrease upon decreasing $t$ as antiferromagnetic 
coupling gets more
effective and canting consequently harder.
As $t$ is further lowered, the anisotropy starts to play a role,
and most of the spins antialign in the easy plane: 
therefore, compared to the isotropic case, 
the number of spins responding to a field applied
in the plane (along $z$) decreases (increases). 
As a consequence, $\chi_{\rm u}^{xx}$ decreases faster, while 
$\chi_{\rm u}^{zz}$ slows down its decrease;
the further reduction of out-of-plane 
fluctuations is eventually responsible for the low-temperature 
increase of $\chi_{\rm u}^{zz}$. 
A clear minimum consequently appears and 
marks the crossover to XY
behaviour at the temperature $t_{\rm co}$,
where the out-of-plane component of 
the antiferromagnetic coupling becomes irrelevant.
No particular feature is instead seen at the transition.

In order to check the direct relation between the crossover 
phenomenon and the out-of-plane fluctuations, 
we consider our data relative to finite-size out-of-plane 
staggered magnetization, $m^z_{\rm s,L}$ ; 
such quantity, which obviously vanishes at all temperatures in 
the thermodynamic limit, is quite useful when local spin 
configurations are under analysis, as in our case.
In the inset of Fig.~\ref{f.chiU} we show data for different 
lattice sizes and $\Delta = 0.02$. By lowering $t$, 
$m^z_{\rm s,L}$ increases (as expected in the isotropic 
behaviour) above $t_{\rm co}$, and 
decreases (as expected in XY behaviour) below $t_{\rm 
co}$: a stable maximum is seen at $t_{\rm co}$.  
From the above evidences we obtain  
the estimates $t_{\rm co}(\Delta=0.02)=0.30(1)$ 
and $t_{\rm co}(\Delta=0.001)=0.225(10)$.
Signatures of the crossover are also present 
in the staggered 
out-of-plane susceptibility and out-of-plane correlation
length (shown in Ref. \onlinecite{CRTVVprb02}),
displaying a maximum at a temperature quite close
to $t_{\rm co}$. 

\begin{figure}
\includegraphics[bbllx=26pt,bblly=60pt,bburx=516pt,bbury=430pt,%
     width=65mm,angle=0]{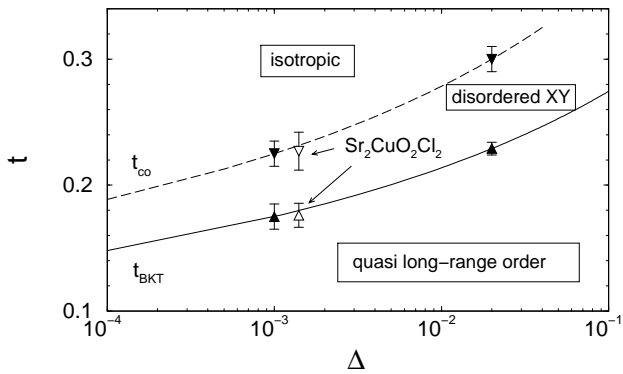}
 \caption{\label{f.phdiagr} Phase diagram of the 
weakly EP model on the square
lattice. Down triangles: $t_{\rm co}$; 
up triangles: $t_{_{\rm BKT}}$. Solid and dashed lines
are logarithmic fits. Also reported are the 
transition and crossover temperatures of 
Sr$_2$CuO$_2$Cl$_2$ (see text).}
\end{figure}

The above results suggest that the crossover phenomenon 
is peculiar to the system and is due to the suppression of 
out-of-plane fluctuations. 
In particular, $t_{\rm co}$ is the 
temperature below which the weakly 
anisotropic system behaves like a planar rotator model
with a spin length 
effectively reduced by out-of-plane fluctuations.
Other authors have referred to such crossover as due 
to a "spin-dimensionality reduction", meaning 
the loss of one spin component~\cite{Vakninetal97}.

By the semiclassical reasoning in Appendix B of Ref.\onlinecite{CRTVVprb02} 
we can predict $t_{\rm co}$ to depend on the anisotropy parameter 
$\Delta$ according to
\begin{equation}
t_{\rm co}\simeq \frac{4\pi\rho_{_{\rm S}}/J}{\ln(C/\Delta)}~~,
\label{e.tcoDelta}
\end{equation}
where $\rho_{_{\rm S}}$ is the renormalized spin stiffness of the 
quantum isotropic model, and $C$ is a constant.
A logarithmic fit to our data, shown in Fig. \ref{f.phdiagr},
gives $C=160$ and $\rho_{_{\rm S}}=0.214 ~J$, which 
compares well with the known value~\cite{Beard}
$\rho_{_{\rm S}}=0.180 ~J$.

\begin{figure}
\includegraphics[bbllx=26pt,bblly=60pt,bburx=516pt,bbury=430pt,%
     width=65mm,angle=0]{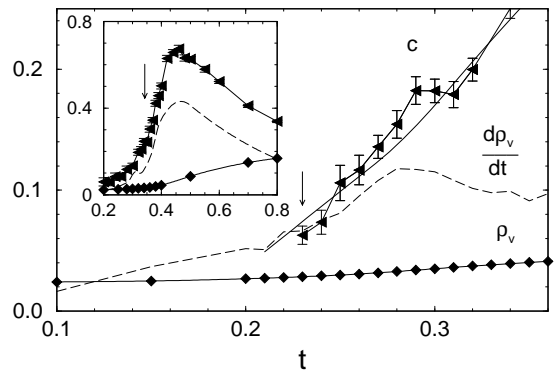}
 \caption{\label{f.sphvortex} Specific heat, vortex density
 and its temperature derivative for the EP model with
 $\Delta = 0.02$. Dashed line: numerical
 derivative of splines for the vortex density data;  
 solid line: specific heat of the isotropic model
 (from a numerical interpolation to the data of 
 Ref.~\onlinecite{KimT98}).
 Inset: same picture for $\Delta = 1$
 (vortex density data from Ref. \onlinecite{Makivic92}).
 When not visible, the error bars are smaller than
 the symbol size; arrows are the BKT critical temperatures.}
\end{figure}

The onset of XY behaviour below $t_{\rm co}$
is further supported by the temperature dependence
of the specific heat $c$.
In the planar rotator model such quantity 
displays a maximum well above the BKT transition. 
Fig.~\ref{f.sphvortex} shows our data for the
weakly anisotropic model with $\Delta=0.02$: an embryonic
peak, out of the isotropic component,  is seen at 
$t = 0.29(1)$, slightly below $t_{\rm co}$. 
To clarify the actual meaning
of such peak, we have also considered the density 
of in-plane vortices with unitary vorticity,
defined as~\cite{Bettsetal81}
\begin{equation}
\rho_{_{\rm V}} = 
\frac{1}{8}
\left\langle 1- 8~\hat{S}^x_{\bm i}\hat{S}^x_{\bm l}
+ 16 ~\hat{S}^x_{\bm i}\hat{S}^y_{\bm j}
\hat{S}^x_{\bm l}\hat{S}^y_{\bm m}\right\rangle~~,
\label{e.rhovortex}
\end{equation}
where $(\bm i, \bm j, \bm l, \bm m)$ defines a plaquette of the
square lattice (indices are ordered counterclockwise).
Choosing $z$ as quantization axis, the above quantity is off-diagonal.
The estimator of the bilinear term in Eq. \eqref{e.rhovortex}
can be found in Ref. \onlinecite{Evertz02}. In the same spirit
we introduce the estimator for the quartic term in the context 
of the loop algorithm~\cite{Roscilde02}.

By considering QMC data for $\Delta = 1$, shown in the inset of
Fig. \ref{f.sphvortex}, we observe that the temperature 
derivative of $\rho_{_{\rm V}}$ displays
a maximum where the specific heat exhibits its peak.
Both features mark therefore the
onset of the formation of V-AV pairs.
In Fig. \ref{f.sphvortex} we show that this picture 
is fully reproduced by our data for $\Delta=0.02$, thus confirming
the XY character of our weakly anisotropic system.

From the experimental point of view,
the above findings may be firstly used to easily characterize 
the anisotropy of the layered antiferromagnets: 
if a minimum in the out-of-plane component of the uniform 
susceptibility is observed above the transition, this
is a signature of EP anisotropy (while
a minimum {\it at} the transition suggests an
easy-axis anisotropy, as shown in Ref. \onlinecite{CRTVVprb02}).  
Once $t_{\rm co}$ has been experimentally determined,
Eq. \eqref{e.tcoDelta} allows one to get an independent estimate of 
the bare anisotropy parameter $\Delta$ of Eq.~\eqref{e.xxzmodel}.
As shown below, this is quite an essential point for
interpreting the experimental data.

We now consider the layered cuprate
Sr$_2$CuO$_2$Cl$_2$, whose intra-layer spin-spin coupling 
of the Cu$^{2+}$ ions has been 
proposed~\cite{Grevenetal95} to be governed by the 
Hamiltonian~(\ref{e.xxzmodel}) with $J=1450$ K.
As for the anisotropy parameter $\Delta$, the experimental 
analysis~\cite{Grevenetal95} 
hands us the renormalized value $\Delta^{\rm exp}=0.00014$, 
extracted from low-temperature measurements of the 
spin gap in the out-of-plane branch of the 2D spin 
waves propagating in the Cu-layers, 
by means of the approximated expression 
$G = 4JSZ_c\sqrt{2\Delta^{\rm exp}}$, where $Z_c$ is the 
zero-temperature value of the spin-wave velocity renormalization 
coefficient of the isotropic Heisenberg system~\cite{Igarashi92}.
The above $\Delta^{\rm exp}$ is not hence the  
bare value appearing in 
Eq.~(\ref{e.xxzmodel}), as it already contains quantum 
renormalizations. In fact, a more refined self-consistent
spin-wave-theory 
calculation~\cite{Roscilde02} leads to the expression 
$G=4JSZ_c\sqrt{2Z_\Delta\Delta}$ with $Z_\Delta=0.099$,
and hence $\Delta=\Delta^{\rm exp}/Z_\Delta\simeq 0.0014$.

\begin{figure}
\includegraphics[bbllx=26pt,bblly=60pt,bburx=516pt,bbury=430pt,%
     width=65mm,angle=0]{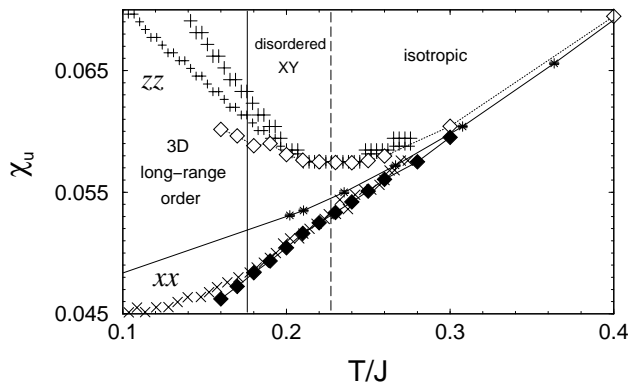}
 \caption{\label{f.SCOC} Uniform susceptibility
of Sr$_2$CuO$_2$Cl$_2$ compared to the theoretical
predictions for the EP model with $\Delta=0.001$.
$\times$'s: $\chi^{xx}_{\rm u}$ of Sr$_2$CuO$_2$Cl$_2$ 
(from Ref. \onlinecite{Vakninetal90}) (with gyromagnetic
factor $g$=2); thin and thick $+$'s: 
$\chi^{zz}_{\rm u}$ of Sr$_2$CuO$_2$Cl$_2$
(from Ref. \onlinecite{Vakninetal97})
with $g$=2  and $g$=2.46, respectively.  
Other symbols and error bars as in Fig. \ref{f.chiU}.}
\end{figure} 

Experimental data relative to the uniform susceptibility 
of Sr$_2$CuO$_2$Cl$_2$  
(from Refs.~\onlinecite{Vakninetal97} and \onlinecite{Vakninetal90})
are shown in Fig. \ref{f.SCOC} together with our results for 
$\Delta=0.001$; 
a constant offset for the experimental data has 
been introduced in order to take into account spurious 
temperature-independent contributions~\cite{Miller02}
to the measured 
values of $\chi_{\rm u}^{\alpha\alpha}$.
The agreement is excellent for both the in-plane and the
out-of-plane component; considering that 
no adjustment has been introduced 
for the temperature axis, the position of 
the minimum of $\chi_{\rm u}^{zz}$ is nicely reproduced, and a
crossover temperature $t_{\rm co}\simeq 0.227$ for the real
compound is determined. 
When the critical region is approached, experimental 
data deviate from the theoretical predictions due to the fact 
that a purely 2D model is no longer sufficient to capture
the thermodynamic behaviour of the real magnet, 
which is in fact characterized by an experimentally 
observed  N\'eel transition at $t_{_{\rm N}}=0.176$.
Such transition temperature compares
remarkably with the BKT transition temperature
of the purely 2D model, $t_{_{\rm BKT}} = 0.175(10)$:
this confirms that the 3D ordering is induced 
by the incipient intra-layer BKT transition. 

From our analysis we conclude that the experimental observation of a
minimum in the transverse uniform susceptibility at a 
temperature which is
well above the critical region, is a signature of EP character
neatly detectable in real compounds even with very weak 
anisotropy. To this respect, we notice that the bare anisotropy
value, needed for a quantitative comparison between 
experimental and theoretical results, significantly differs
from the (measured) renormalized value, due to the relevance
of quantum effects. Finally, two distinct regimes,
separated by the crossover, can
be identified in the disordered phase: 
an {\it isotropic} regime 
for $t > t_{\rm co}$ and a {\it disordered XY} regime 
for $t_{_{\rm N}} < t < t_{\rm co}$. It is 
relevant that the crossover temperature 
can be obtained by measuring non-diverging quantities above the 
critical region, which is an experimentally feasible task.

In summary, by means of quantum Monte Carlo
simulations we have shown that two-dimensional $S=1/2$
square-lattice antiferromagnets with 
a weak easy-plane anisotropy
display a crossover from a high-temperature isotropic
behaviour to a genuinely 2D XY behaviour at a 
temperature $t_{\rm co}$ which stays
around 30$\%
$ above the critical temperature $t_{_{\rm BKT}}$. 
We have moreover shown that  Sr$_2$CuO$_2$Cl$_2$ realizes
these predictions, and therefore stands as a very clean
example of weakly planar quasi-2D antiferromagnet. 
This represent a clear evidence of non-critical
XY behaviour in a real layered antiferromagnet. 
Other evidences for an isotropic-to-XY crossover
in Sr$_2$CuO$_2$Cl$_2$ come from neutron scattering
~\cite{Vakninetal97} and NMR ~\cite{Suhetal96} experiments,
and enforce our conclusions. 
Further work on such crossover effect in
the above compound, as well as in other layered antiferromagnets
with easy-plane anisotropy (as, e.g., Pr$_2$CuO$_4$),
opens the perspective for a systematic 
investigation of the vortex phase in purely magnetic systems. 

We thank D. Vaknin, L. L. Miller and V. Tognetti for fruitful
discussions and correspondence. 
This work has been partially supported by the COFIN2000-MURST fund
and the INFM Advanced Parallel Computing Project.

\end{document}